# Crossed-anisotropy films for magnetic tunnel junctions and magnetic memory applications

A.N. Grigorenko and D.J. Mapps,

*Centre for Research in Information Storage Technology*

*Department of Communication, Electronic and Electrical Engineering,*

*University of Plymouth, Drake Circus, Plymouth, Devon PL4 8AA, United Kingdom.*

**Abstract**

A prototype of magnetoresistive random access memory (MRAM) based on magnetic tunnel junctions (MTJ) was fabricated with crossed-anisotropy of magnetic layers on either side of the tunnelling barrier layer. It is demonstrated that the introduction of crossed-anisotropy results in smaller switching fields and better switching times compared to the conventional case of aligned anisotropies. The magnetoresistive properties of fabricated devices are in good agreement with the micromagnetic model.

# I. Introduction

Recently, the development of magnetic random access memory (MRAM) has been given a strong boost from the rapidly growing technology of spin tunnelling. Spin valves and magnetic tunnel junctions (MTJ) are very promising candidates for high density, fast and comparatively cheap memory [1-5]. They are non-volatile, non-destructive and could operate at low voltage. A large effort has been invested to develop MTJ technology and fabricate successful commercial devices, which has resulted in 1Mbit memory chips [6]. However, the need to improve the basic working characteristics of MTJ for MRAM still exists.

A magnetic tunnel junction consists of two magnetic metal layers separated by a thin dielectric barrier. Usually, the anisotropies of MTJ magnetic layers are aligned parallel along the same direction and the switching field is also applied along this direction. Such geometry brings some disadvantages. It is well known that rotation of magnetisation is of gyroscopic nature and the rotation of magnetization happens in the direction perpendicular to the applied field and that of magnetization. In other words, the torque due to an applied field is proportional to the vector product of magnetic field and magnetisation. Thus, in the usual case where the magnetic field and anisotropies are perfectly aligned, the rotation of the magnetisation of a free layer can not start, either with the applied field torque (which is zero) or the effective relaxation field (which is also zero in a state of a stored information bit). As a result, the switching process needs magnetic fluctuations or incomplete relaxation in order to commence, see (5) below. This means that higher fields are required to jump over the maximum of anisotropy energy (to write an information bit) and the switching process is slower than the gyroscopic motion due to the applied magnetic field. The natural way to use the magnetic anisotropies and the applied field torque is to deliberately misalign the direction of the driving field and the anisotropies of MTJ layers.

In the paper we show both experimentally and theoretically that introducing the crossed-anisotropy of the magnetic layers into a MTJ structure favours smaller switching fields (currents) and may improve the switching time. Since magnetic tunnel junctions with crossed-anisotropy do not rely entirely upon the fluctuations and relaxation processes they should have much better stability of switching characteristics and provide control over the switching time. It is easy to check that fields of about 100 Oe are required to achieve 1 ns switching time in microchips with the gyroscopic ratio of $10^7$ 1/(Oe·sec). As a result, high values of switching fields and driving currents are needed in the conventional case of aligned anisotropies. A MTJ with crossed-anisotropy uses the anisotropy fields from the very beginning of the switching process (in addition to the driving field). Since the switching process starts at smaller fields in MTJ with crossed-anisotropy, smaller driving currents can be used to read/write information from/to MRAM chips. This could result in a decrease of power consumption by memory chips. The decrease of magnetic switching time and much better homogeneity of switching characteristics is also very favourable for the development of a fast memory. Because these changes happen due to a simple additional step in the MTJ fabrication, we believe that crossed-anisotropy is a very promising method to be used with MRAM devices.

**II. Fabrication and Results on Switching Fields.**

The 2×2 and 8×8 bits prototypes of a MRAM memory chip with MTJ were fabricated on a silicon substrate using RF diode sputtering in a Nordico NM2000 system with a base pressure better than $10^{-7}$ Torr. Several different MTJ structures have been studied: i) Ta 5 nm/NiFe 20-40 nm/AlO 1-2 nm/Co 20-40 nm, ii) Ta 5 nm/NiFe 20-40 nm/CoFe 5-10nm/AlO 1-2 nm/CoFe 30-40 nm, iii) Cr 5nm/Co 20-40 nm/AlO 1-2 nm/NiFe 20-40 nm. The whole structure also included bottom and upper leads which were used as reading and writing lines. The deposition of the Al layer was performed on a pre-oxidised magnetic layer produced by natural oxidation in 210 mTorr of pure $O_2$ inside the chamber for 4-6 hours. Pre-oxidation of a magnetic layer greatly improved the homogeneity of the deposited nanolayer of Al and reduced the number of "holes" inside the tunnelling layer, see, e.g., [7]. We believe that a strong chemical bond between aluminium and oxygen is the reason for this improvement. Natural oxidation of the bottom magnetic layer is a homogeneous process which provides a very thin homogeneous layer of NiFe (CoFe) oxide on the top of the bottom layer. The magnetic atoms of the oxide are replaced by aluminium atoms during a subsequent deposition of Al because of the higher chemical activity of Al. Strong bonds between Al and O ensure that the deposited Al layer on the top of the oxide layer is more homogeneous than that of Al on the bare bottom magnetic layer. The Ta layer ensures that the microstructure of the NiFe has a predominant (111) plane parallel to the surface of the film [8].

Natural in-chamber oxidation for 14-24 hours in 210-320 mTorr of pure oxygen was also used to produce a thin dielectric barrier of alumina after the Al deposition. The anisotropy in the magnetic layers was induced by permanent magnets placed into a substrate holder and/or by subsequent annealing in magnetic field. The deposition process was interrupted after deposition of the first magnetic layer so that the sample could be taken out and repositioned at a different angle inside the sample holder in order to produce MTJ with crossed anisotropy. The result was a substantial decrease in MTJ resistance with the final devices never showing a MTJ effect more than 5%. A much better way to introduce crossed-anisotropy into MTJ's is to make use of a sample holder with a re-positionable magnet or utilising pinning layers of (artificial) antiferromagnetic combined with annealing in magnetic fields of different directions. However, the fabricated MRAM memory chips were sufficiently good to demonstrate the main claims of this paper.

The bottom and upper electrodes were patterned using optical lithography. MTJ were produced by a lift-off technique in order to minimise the contact of fabricated junctions with the atmosphere and exclude completely the contact with the resist developer and rinsing agent (water) which had the effect of etching the alumina. Figure 1 shows a typical (a) 2x2 memory element and (b) 3x3 piece of a 8x8 MRAM chip. The junction sizes span a range from 1 μm to 200 μm. Figure 1(c) shows the alignment of the directions of magnetic anisotropies and the applied field. The direction of the bottom layer anisotropy field $H_{a1}$ is taken as the reference direction, the direction of the upper layer anisotropy field $H_{a2}$ makes an angle $\theta$ with the reference direction and the magnetic field $H(t)$ is applied under an angle $\varphi$ with respect to the reference direction.

Now we turn to the discussion of the influence of crossed-anisotropy on the switching fields of magnetic layers measured from magnetization loops. Figure 2 shows three typical curves of magnetoresistance of magnetic tunnelling junctions made with different degrees of crossed-anisotropy. For simplicity, only a half of the magnetisation cycle is plotted (with the other half being symmetric along y-axis). The magnetoresistanse of the MTJ structure of Ta 5 nm/NiFe 20-40 nm/AlO 1-2 nm/Co 20-40 nm is chosen as being representative. The results for all other structures were qualitatively the same. Figure 2(a) depicts the conventional case of aligned anisotropies and magnetic field, with a typical 5-7 Oe of the switching field for NiFe and about 12-14 Oe of the switching field for Co. We found that a MTJ structure with crossed-anisotropy made from the same magnetic materials and under the same fabrication conditions has smaller switching fields (measured from the magnetization loops) for both magnetic layers. Figure 2(b) demonstrates this for the case of extreme crossed-anisotropy of $\theta = 90°$ and magnetic field applied in the direction of $\varphi = 45°$ (MTJ size is 20 μm). One can easily see that the switching field for the NiFe layer dropped to 1-2 Oe, while the switching field for the Co layer dropped to about 5 Oe. All junctions with crossed-anisotropy showed a decrease of the switching fields. This effect was the same for junctions of different sizes. The best results were obtained for the case of a crossed-anisotropy angle of $\theta = 45°$ with the magnetic field applied at the angle of $\varphi = 20°$. This case is shown in Fig. 2(c) for the same NiFe/AlO/Co structure where the switching field of the NiFe layer is about 1 Oe and the switching field for the Co layer is about 2-3 Oe.

**III. Micromagnetic Model.**

The whole variety of the magnetoresistance curves for the junctions with different crossed-anisotropy is described well by the micromagnetic model of the homogeneous rotation of the magnetisations $\vec{M}_1$ and $\vec{M}_2$ of two coupled magnetic layers. The total energy of two layers is written as

$$E = E_a + E_c + E_H. \tag{1}$$

Here $E_a$ is the anisotropy energy

$$E_a = K_1 (M_{1z})^2 + K_2 (M_{2z})^2 - \frac{H_{a1}}{2|\vec{M}_1|} (\vec{M}_1 \vec{o}_{a1})^2 - \frac{H_{a2}}{2|\vec{M}_2|} (\vec{M}_2 \vec{o}_{a2})^2, \qquad (2)$$

$K_1$ and $K_2$ are the hard axis anisotropies, $\vec{o}_{a1}$ and $\vec{o}_{a2}$ are the directions of the in-plane anisotropies, $H_{a1}$ and $H_{a2}$ are the in-plane anisotropies, $E_c$ is the coupling energy of two layers

$$E_c = -J(\vec{M}_1 \vec{M}_2), \qquad (3)$$

where $J$ is the coupling density and $E_H$ is the magnetization energy in the external magnetic field $H(t)$

$$E_H = -\vec{H}(t)\vec{M}_1 - \vec{H}(t)\vec{M}_2. \qquad (4)$$

The motion of magnetization was taken in Landau-Lifshitz form

$$\frac{d\vec{M}}{dt} = -\gamma \vec{M} \times \vec{H}_{eff} - \lambda \vec{M} \times (\vec{M} \times \vec{H}_{eff}), \qquad (5)$$

where $\vec{H}_{eff} = -\partial E / \partial \vec{M}$ is the effective field and $\lambda$ is the Landau parameter ($\lambda = \gamma\alpha / |\vec{M}|$, $\alpha$ is the dimensionless dissipation parameter).

Two coupled equations (5) for both magnetizations were solved numerically by the Runge-Kutta method with an adaptive step. We found that both magnetizations prefer to stay inside the easy plane and the values of the magnetization perpendicular to the surface of the MTJ were small throughout all dynamics of magnetization. This allowed us to use an in-plane simplification of the equations (5) with the anisotropy energy

$$E_a = -\frac{H_{a1} M_1}{2} \cos^2(\phi_1) - \frac{H_{a2} M_2}{2} \cos^2(\phi_2 - \theta), \qquad (6)$$

where $\phi_1$ and $\phi_2$ are angles of magnetizations $M_1 = |\vec{M}_1|$ and $M_2 = |\vec{M}_2|$ with respect to the reference direction. These angles were the dynamic variables of the problem. The coupling energy is then

$$E_c = -JM_1 M_2 \cos(\phi_1 - \phi_2) \qquad (7)$$

and the Zeeman energy is

$$E_H = -H(t)(M_1 \cos(\phi_1 - \varphi) + M_2 \cos(\phi_2 - \varphi)). \qquad (8)$$

The in-plane model requires much less computation time and (for the static magnetization curves and switching fields) yields results which deviate less than 1% compared with the solutions of the exact equations (5). For the switching times calculations the exact equations have been used.

The main parameters of the model (1)-(8) were found from different experiments (magnetic anisotropies were evaluated from the magnetization curves and magnetizations were measured using a vibrating sample magnetometer). The Landau-Liftshitz-Gilbert dimensionless constant $\alpha$ constant was taken as $\alpha = 0.1$. For the coupling energy between two layers we have used Neel "orange peel" evaluation [1] (where the coupling is attributed to magnetostatic energy arising due to the interface roughness). Figure 3 shows the calculated magnetoresistance curves of MTJ with a different degree of crossed-anisotropy. One can see that the calculated curves are in a good qualitative agreement with those depicted in Fig. 2 and indeed show a considerable decrease of the switching fields of a magnetization loop for a crossed-anisotropic MTJ. (Note that a small misalignment of the applied magnetic field $\varphi = 7°$ was introduced in the conventional case in order to break the degeneracy of the calculations.)

We found this qualitative agreement of computer simulations with the observed behaviour of MTJ in all cases of crossed-anisotropy and in the presence of additional constant fields, e.g., Fig. 4(a) displays an asymmetric tunnelling magnetoresistance curve measured in the presence of an additional magnetic field of 0.9 Oe applied at angle of 7° to the reference direction. The tunnelling magnetoresistance magnetization curve calculated with the help of (1)-(8) describes well the experimental measurements, see Fig. 4(b). Figure 4(a)-(b) reveals that tunnelling magnetoresistance of MTJ with crossed-anisotropy is very sensitive to additional applied constant fields. Another interesting result observed in fabricated MTJ was the fast and reproducible changes of the tunnelling magnetoresistance at some particular values of crossed-anisotropy and applied fields. Figure 4(c) (the experimental dependence) and (d) (the calculated curve) demonstrates this tendency for the case of the extreme crossed-anisotropy $\theta = 90°$ with the magnetic field applied at the angle of $\varphi = 70°$. High magnetic susceptibility as well as strong dependence of the magnetization curves on additional magnetic fields can be used in the development of magnetic field sensors based on MTJ. We shall describe the application of MTJ with crossed-anisotropy in magnetic field sensors elsewhere. We note again that the calculated changes in magnetization are in good qualitative agreement with the measured data.

**IV. Calculation of Switching Time.**

The decrease of the switching field for MTJ with crossed-anisotropy can be used to reduce the switching currents and hence to decrease power consumption of memory chips. In other arrangements, one may improve the switching times of the microchips by keeping the switching currents constant and increasing the anisotropies of the magnetic layers. Remarkably, the switching times of the MTJ structure with crossed-anisotropy are better than that of the conventional case even when all anisotropies are kept at the same level. This improvement can be anticipated because the deliberate misalignment of magnetic field and anisotropies makes possible to use the relaxation due to anisotropy fields from the very beginning of the switching process in addition to the applied magnetic field torque. We demonstrate this result using computer modelling.

Figure 5(a) shows the calculated MTJ responses when the applied magnetic field of 2.4 Oe (the dotted line) suddenly reverses its direction in the conventional case of aligned anisotropies (dashed line) and in the case of crossed-anisotropy of $\theta = 45°$ with the magnetic field applied at the angle of $\varphi = 25°$ (solid line). The reversal of the field happens at the 20 ns point. The time response of the crossed-anisotropic MTJ is about 2.5 ns while the response of the conventional MTJ is about 7.5 ns at the same magnitude of layer anisotropies (the magnetic field was applied at the angle of 3° in respect with the anisotropy direction in order to start the switching process). This yields about a three-fold decrease of the switching time due to the crossed-anisotropy.

Note that the switching process of the crossed-anisotropic MTJ starts immediately after an application of the driving impulse (see the 20 ns point on the time scale), whilst in the conventional case the switching process kicks in only after a long time delay of 5 ns after the impulse. The long time delay or "incubation" time of the magnetoresistance response is due to a very small effective driving field for the magnetization dynamics in the case of aligned anisotropies and magnetizations. In the conventional case the magnetization reversal relies on fluctuations and/or small misalignment of the magnetic field which effects in high fluctuations of the switching time. As a result, it is difficult to achieve a control (homogeneity) of switching characteristics over a large number of MTJ cells. For example, the incubation time of the switching process shown in Fig. 4 increases by about 1.5 times for the magnetic field misalignment of 2° and does not happen at the same level of field at misalignment of 1°. MTJ with crossed-anisotropy are free from this disadvantage and show consistent switching characteristics in the case of small variations of all parameters (anisotropies and fields).

Another important feature of the switching process based on the discussed properties of the magnetization reversal is that a MTJ with crossed-anisotropy can be switched by a short impulse of smaller amplitude (at a fixed impulse duration) compared to a conventional MTJ. Figure 5(b) demonstrates that a 10 ns impulse of the magnetic field of 2.4 Oe easily switches magnetization in a MTJ with crossed-anisotropy of $\theta = 45°$ (the field is applied at $\varphi = 25°$), see the solid line that represents the calculated tunnel magnetoresistance. On the contrary, the incubation time makes impossible to switch magnetization of the free layer in the conventional MTJ with the same level of magnetic anisotropies, see the dashed line. This observation could be important when the necessity to reduce switching times of MRAM below the nanosecond limit will arise.

The fabricated memory cells (with and without crossed-anisotropy) were easily switched by 10 ns pulses of magnetic field. The measurements confirmed that the crossed-anisotropic MTJ requires smaller amplitude of a magnetic field impulse of a fixed duration in order to change a cell state (to write information) than MTJ without crossed-anisotropy in accordance with the discussion above. Experiments on nanosecond switching of fabricated memory chips are now being carried out and will be reported in a later paper.

**V. Conclusions.**

To summarise, we propose to use crossed-anisotropy of magnetic layers in magnetic tunnel junctions for MRAM memory based on MTJ in order to improve the switching characteristics of memory cells and to homogenise MTJ working parameters. We have shown theoretically that the switching field and time decrease and become more consistent over a large number of memory cells. The 2x2 and 8x8 prototypes of memory chips were fabricated and tested. The measured tunnel magnetoresistance of these chips is described well by the micromagnetic theory of the magnetization rotation. The crossed-anisotropic MTJ have demonstrated a considerable decrease of switching fields (in about 3 times) in comparison with the conventional case of aligned anisotropies.

**Acknowledgments.**

Authors would like to thank Mark Blamire and Amanda Petford-Long for useful conversations, Nick Fry and Phil Brown for the technical assistance. This work has been supported by the EPSRC Grant No. GR/M23977/01.

**Figure Captions.**

Fig. 1.

Ptototypes of MRAM memory based on MTJ: (a) 2x2 bit chip, (b) 3x3 fragment of 8x8 bit chip, (c) mutual alignment of MTJ anisotropies and the applied magnetic field.

Fig. 2.

Measured tunnel magnetoresistance during magnetization of MTJ with a different degree of crossed-anisotropy: (a) the conventional case of aligned anisotropies, (b) the extreme crossed-anisotropy of 90°, field is applied at an angle of 45°, (c) the crossed-anisotropy of 45°, the magnetic field is applied at 25°. The size of measured junctions is 20 μm.

Fig. 3.

Tunnel magnetoresistance during magnetization of MTJ with a different degree of crossed-anisotropy calculated with the help of the micromagnetic model: (a) the conventional case of aligned anisotropies, (b) the extreme crossed-anisotropy of 90°, field is applied at an angle of 45°, (c) the crossed-anisotropy of 45°, the magnetic field is applied at 25°.

Fig. 4.

Tunnel magnetoresistance during magnetization: (a) the extreme crossed-anisotropy of 90°, field is applied at an angle of 25° in the presence of an additional constant magnetic field of 0.9 Oe at an angle of 7°, (b) the micromagnetic simulations for the experimental result shown in (a), (c) the extreme crossed-anisotropy of 90°, field is applied at an angle of 70°, (d) the micromagnetic simulations for the experimental result shown in (c).

Fig. 5

Switching process in MTJ with and without crossed anisotropy: (a) an abrupt change of the field from - 2.4 to 2.4 Oe, (b) an impulse of the "writing" field of 2.4 Oe. The dotted line – magnetic field, the dashed line – MTJ without crossed-anisotropy, the solid line – MTJ with crossed-anisotropy of 45°, the field is applied at an angle of 25°.

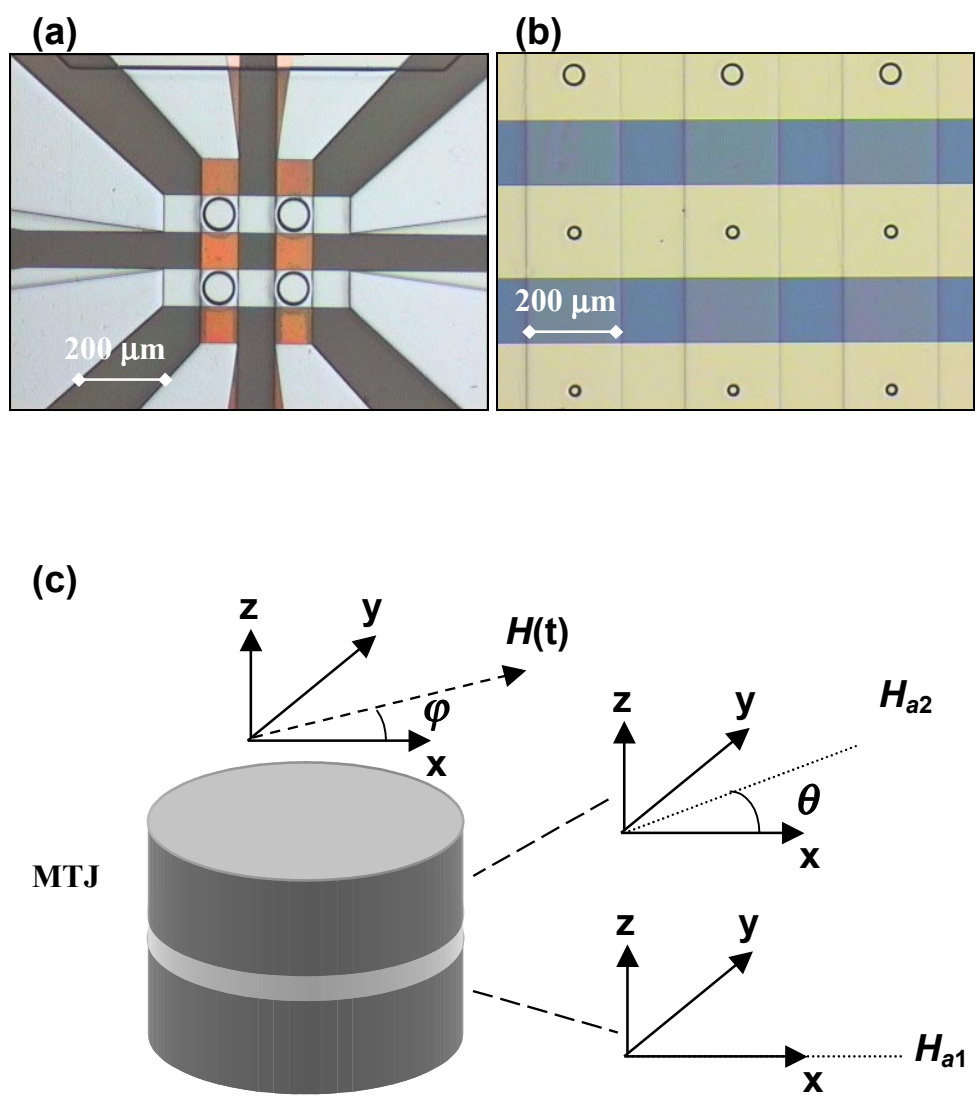

Fig. 1.

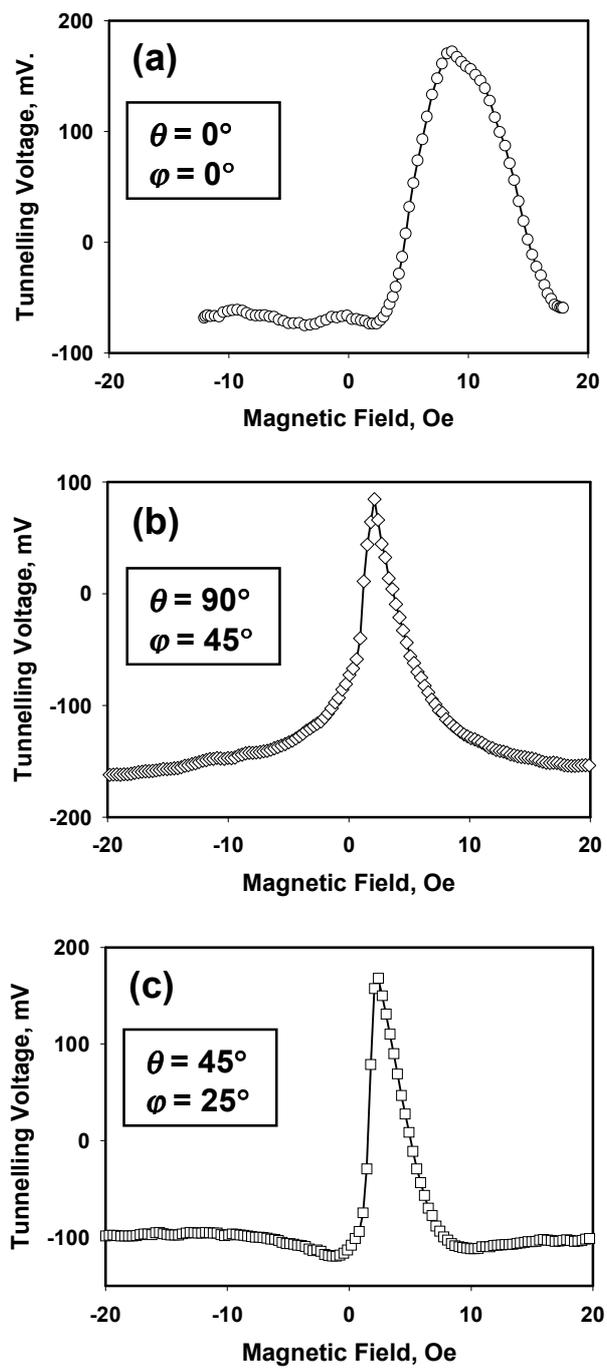

Fig. 2.

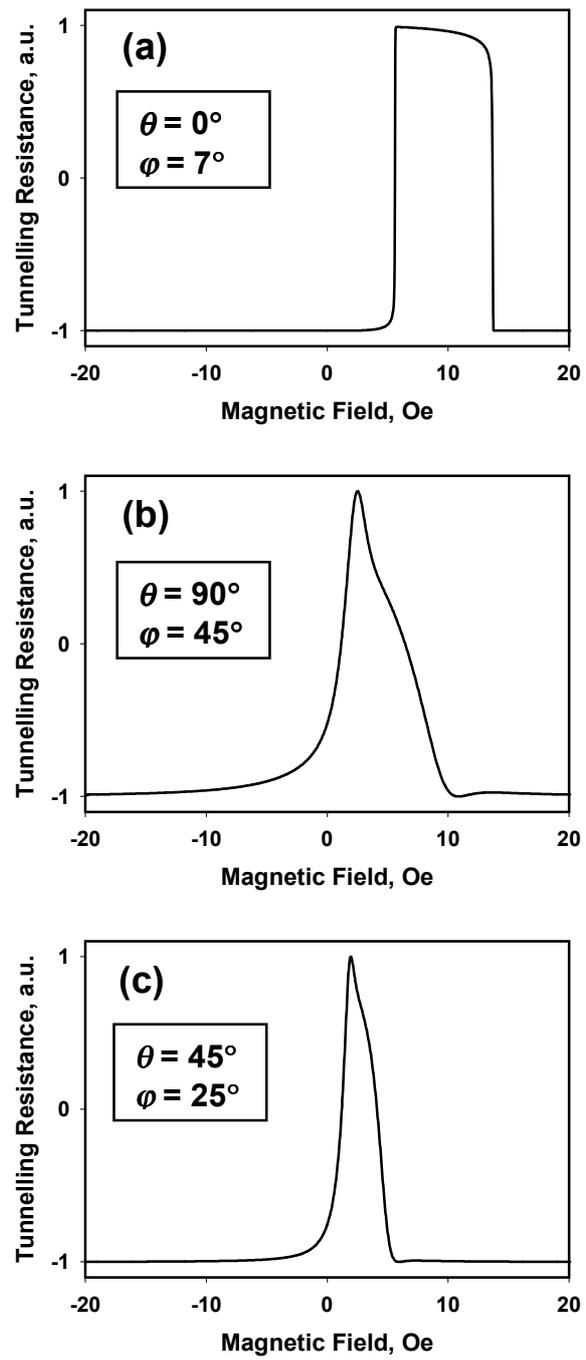

Fig. 3.

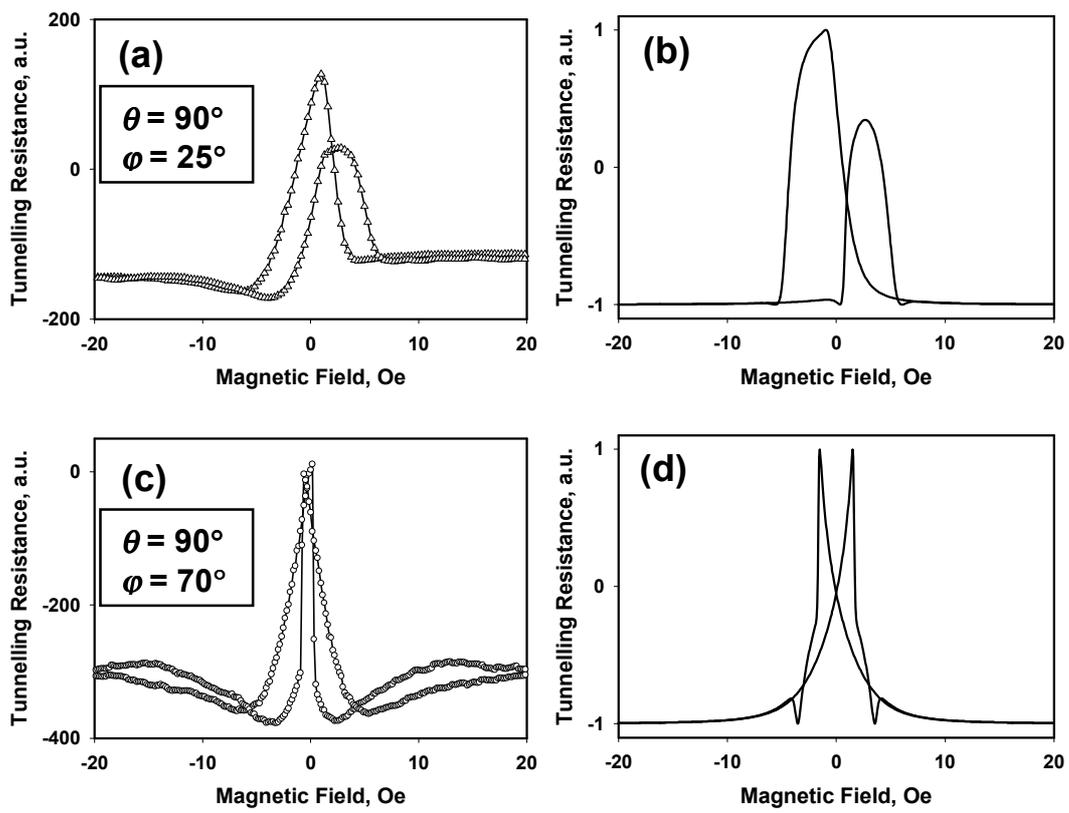

Fig. 4.

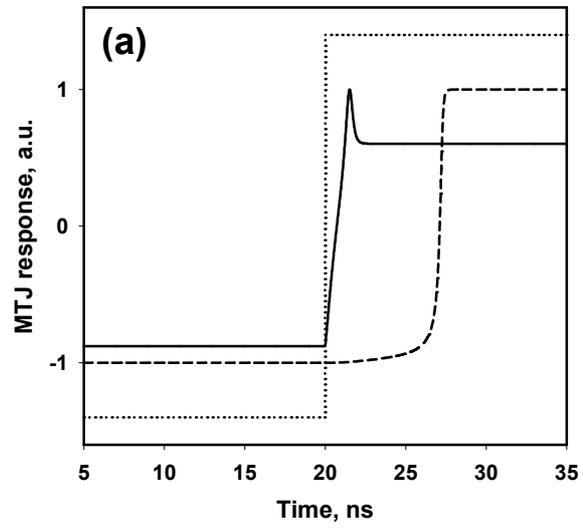

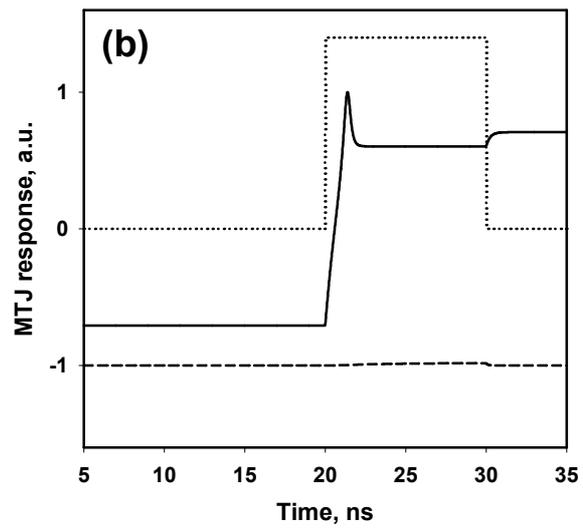

Fig. 5.